\documentclass[twocolumn,showpacs,preprintnumbers,amsmath,amssymb,superscriptaddress]{revtex4}

\usepackage{graphicx}
\usepackage{bm}

\begin{document}


\title{Coherent single electron spin control in a slanting Zeeman
field}

\author{Yasuhiro Tokura}%
\affiliation{NTT Basic Research Laboratories, NTT Corporation, Atsugi-shi,
Kanagawa 243-0198, Japan
}%
\affiliation{%
Tarucha Quantum Spin Information Project, ICORP,
JST,
Atsugi-shi, Kanagawa 243-0198, Japan
}%
\author{Wilfred G. van der Wiel}
\affiliation{%
SRO NanoElectronics, MESA$^+$ Institute for NanoTechnology,
University of Twente, P$.$O$.$ Box 217, 7500 AE Enschede, The
Netherlands; PRESTO-JST, University of Tokyo, Hongo, Bunkyo-ku,
Tokyo 113-8656, Japan
}%
\author{Toshiaki Obata}%
\affiliation{%
Tarucha Quantum Spin Information Project, ICORP,
JST,
Atsugi-shi, Kanagawa 243-0198, Japan
}%
\author{Seigo Tarucha}
\affiliation{%
Tarucha Quantum Spin Information Project, ICORP,
JST,
Atsugi-shi, Kanagawa 243-0198, Japan
}%
\affiliation{%
Department of Applied Physics, University of Tokyo,
Hongo, Bunkyo-ku, Tokyo 113-0033, Japan
}%

\begin{abstract}
%
%
We consider a single electron in a 1D quantum dot with a static
slanting Zeeman field. By combining the spin and orbital degrees of
freedom of the electron, an effective quantum two-level (qubit)
system is defined. This pseudo-spin can be coherently manipulated by
the voltage applied to the gate electrodes, without the need for an
external time-dependent magnetic field or spin-orbit coupling.
Single qubit rotations and the C-NOT operation can be realized.
We estimated relaxation ($T_1$) and coherence ($T_{2}$) times, and the (tunable)
quality factor. This scheme implies important
experimental advantages for single electron spin control.
\end{abstract}


\maketitle

\begin{figure}
\includegraphics[width=8cm]{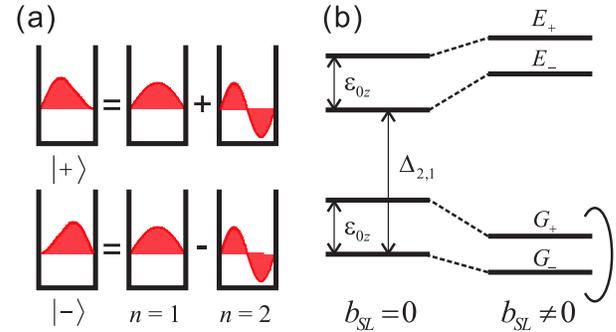}
\caption{\label{fig:level} (a) Schematic representation of how a
spatial oscillation between wavefunctions $|+\rangle$ and
$|-\rangle$ involves hybridization of multiple orbital states. (b)
Energy spectrum of a quantum dot (QD) with two orbital levels (level
spacing $\Delta_{2,1}$) and constant Zeeman energy
$\varepsilon_{0z}$ with/without a magnetic field gradient $b_{SL}$.
The lowest levels, $|G_{\pm}\rangle$, constitute a qubit.
$|E_{\pm}\rangle$ are excited states.}
\end{figure}

\indent Stimulated by electron-spin-based proposals for quantum
computation \cite{loss1998}, a growing interest has emerged in
realizing the coherent manipulation of a single electron spin in a
solid-state environment. The application of the electron's spin --
rather than its charge -- as a quantum bit (qubit) is motivated by
its potentially long coherence time in solids and the fact that it
comprises a natural two-level system. Single electron spin resonance
(SESR) plays a key role in realizing electron-spin-qubit rotation.
Importantly, SESR is also the prime tool for determining the single
electron spin coherence time $T_2$ in confined solid-state systems
such as quantum dots (QDs). The induced Rabi oscillations can be
read out via electron transport \cite{engel0102} or optically
\cite{gywat2004}, giving an estimate for $T_2$. SESR was detected in
paramagnetic defects in silicon \cite{xiao2004} and for nitrogen
vacancies in diamond \cite{jelezko2004a}, but not in semiconductor
QDs so far. Realizing SESR in QDs is hard, not least because of the
necessary high-frequency ($\sim$10 GHz) magnetic field in a
cryogenic ($\sim$100 mK) setup. Waveguides and microwave cavities as
used in conventional ESR \cite{dobers1988}, cause serious heating,
limiting the operation temperature to $\sim$1 K. On-going work in
our group focuses on generating ac magnetic fields
by an on-chip microscopic coil \cite{kodera2004}.\\
\indent In this Letter, we propose a new SESR scheme that eliminates
the need for an externally applied ac magnetic field, and with the
potential of very high and tunable quality factors. An ac voltage is
applied to let an electron in a QD oscillate under a {\it static}
slanting Zeeman field. This effectively provides the electron spin
with the necessary time-dependent magnetic field. Note the analogy
with the Stern-Gerlach experiment, where the spin and orbital
degrees of freedom are coupled by employing an inhomogenous magnetic
field. The spatial oscillation of the electron within the QD
involves the hybridization of orbital states, as schematically
depicted in Fig$.$~\ref{fig:level}a for the case of the two lowest
orbital states, $n=1,2$.
\begin{figure}
\includegraphics[width=8cm]{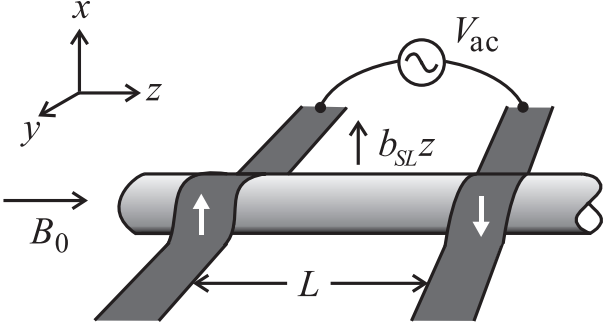}
\caption{\label{fig:dot} Model of the 1D QD in a slanting Zeeman
field. Ferromagnetic gate electrodes (dark grey) are located at
either end of the dot and are magnetically polarized in the
plus/minus $x$-direction, creating a magnetic field gradient
$b_{SL}$. A uniform magnetic field $B_0$ is applied in the
$z$-direction. The spin in the dot is controlled by applying an
oscillating voltage $V_{ac}$ between the two gates.}
\end{figure}
Charge qubits based on double QDs \cite{Wiel03} offer great
tunability, but suffer from short coherence times ($\sim$1 ns)
\cite{hayashi}. Spin qubits on the contrary, enable long coherence
times($\sim$1 $\mu$s) \cite{Petta05}, but are much harder to
control, as pointed out above. Here, we present a hybrid charge-spin
system that is promising both in terms of tunability and coherence.
Analogously, the combination of the flux and charge degrees of
freedom has proved to be fruitful in superconducting qubits
\cite{vion}. We stress that in our system spin-orbit (SO) coupling
is not required, as opposed to earlier work on electron spin control
based on $g$-tensor modulation \cite{kato2003}, and on electric
fields \cite{rashba}.\\
\indent A possible realization of the system is presented in
Fig$.$~\ref{fig:dot}. A 1D conductor like a carbon nanotube or
semiconductor nanowire is gated by ferromagnetic electrodes that
define both the tunnel barriers of the QD and the slanting magnetic
field. Alternatively, the slanting Zeeman field could be provided by
a static inhomogeneity in the nuclear spin polarization or in the
$g$-factor. The total magnetic field is given by $ \bm{B}=b_{SL}
z\bm{i}_x+(B_0+b_{SL} x)\bm{i}_z$, where $B_0$ is the external
uniform magnetic field parallel to the $z$-axis and $b_{SL}$ is the
$z$-direction gradient of the field parallel to the $x$-axis (the
middle of the QD corresponds to $z=0$.). A true 1D system is assumed
with an electron strongly confined in the $x$ and $y$ directions.
Therefore, the inhomogeneous term along the $z$ axis, $b_{SL} x
\bm{i}_z$, can be eliminated (which is there to let $\bm{B}$ obey
Maxwell's equations). \\
\indent The Hamiltonian is $H_0=H_{00}+H_{0s}$, where $
H_{00}=\frac{p^2_z}{2m}+V(z)-g\mu_B B_0 S_z$, and $H_{0s}=-g\mu_B
b_{SL} z S_x$, with $\bm{S}=\frac{1}{2}\bm{\sigma}$, $\bm{\sigma}$
the Pauli spin matrices, $g$ the effective $g$-factor, and $\mu_B$
the Bohr magneton \cite{levitov}. $V(z)$ is the confinement
potential of the QD with length $L$.
The eigenvalues of $H_{00}$ are $
\varepsilon_{n\sigma}=\varepsilon_n+\frac{1}{2}\varepsilon_{0z}
\sigma$, and the eigenfunctions $\langle z|n,\sigma\rangle =
\xi_\sigma \phi_n(z)$, where $n=1,2,\ldots$, $\sigma=\pm 1$ and
$\xi_\sigma$ is the spinor. We define the Zeeman energy
$\varepsilon_{0z}=|g\mu_B B_0|$, which is assumed to be smaller than
the orbital energy level separation
$\Delta_{n,m}=\varepsilon_n-\varepsilon_m\ll U$, with $U$ the
charging energy. The non-zero matrix elements of $H_{0s}$ are
$\langle m, -\sigma|H_{0s}|n,\sigma\rangle \equiv\frac{1}{2}
M_{m,n}$, with a coupling energy $M_{m,n}\equiv E_{SL}
\Upsilon_{m,n}$ where $E_{SL}\equiv -g\mu_B b_{SL} L$ characterizing
the strength of the slanting field, and the form factor
$\Upsilon_{m,n}\equiv \int dz \phi_m^*(z)\frac{z}{L}\phi_n(z)$.\\
\indent By requiring the confining potential to have a mirror
symmetry, i$.$e$.$ $V(z)=V(-z)$, the diagonal coupling energy elements vanish,
namely $M_{n,n}=0$. We employ perturbation theory up to the second
order in $E_{SL}$ and obtain the ground-state energy for a
pseudo-spin $\sigma$, $
G_\sigma=\varepsilon_1+\frac{1}{2}\varepsilon_{0z}\sigma
-\frac{1}{4}\sum_l\frac{M_{1,2l}^2}{\Delta_{2l,1}-\varepsilon_{0z}\sigma},
$ and its wavefunction $
|G_\sigma\rangle=C^{(0)}_\sigma|1,\sigma\rangle +\sum_{l>0}
C^{(1)}_{l\sigma}|2l,-\sigma\rangle
+\sum_{n>0}C^{(2)}_{n\sigma}|2n+1,\sigma\rangle$
\cite{coefficients}. Since we assumed
$\Delta_{2,1}>\varepsilon_{0z}$, the two lowest energy states
$|G_+\rangle$ and $|G_-\rangle$ represent an energetically isolated
qubit (see Fig$.$ \ref{fig:level}). We can disregard higher energy
states, such as $|E_{\pm}\rangle$. For a rectangular confining
potential, we find for the form factor
$\Upsilon_{2n+1,2l}=-\frac{8}{\pi^2}(-1)^{l+n}\frac{2l(1+2n)}{((1+2n)^2-4l^2)^2}$,
while for a harmonic potential $V(z)=\frac{m\omega_0^2}{2}z^2$, we
have
$\Upsilon_{2n+1,2l}=\delta_{n+1,l}\sqrt{n+\frac{1}{2}}+\delta_{n,l}\sqrt{n}$,
where we set $L=\sqrt{\frac{\hbar}{m\omega_0}}$. Therefore,
$\Upsilon_{nm}$ is negligible for large $|n-m|$, and we only
consider $M_{1,2}$ and $M_{2,3}$, which is exact for the harmonic
potential. We define the effective Zeeman energy
$\varepsilon_z\equiv G_+-G_-\sim
\varepsilon_{0z}[1-\frac{1}{2}\frac{M_{1,2}^2}{\Delta_{2,1}^2-\varepsilon_{0z}^2}]$.

We consider the qubit rotation induced by an ac electric field. The
time-dependent perturbation $H_1(t)=eV_{ac}(t)\cdot \frac{z}{L}$ is
applied to the system by introducing an oscillating signal
$V_{ac}(t)=V_0 f(t)$ to the gate electrodes, as shown in
Fig$.$~\ref{fig:dot}. Since $H_1(t)$ is an odd function of $z$ and
is independent of spin, only the off-diagonal matrix elements of
$H_1$ remain, $ \langle G_\sigma|H_1|G_{-\sigma}\rangle
\sim(C^{(1)}_{1\sigma}+C^{(1)}_{1-\sigma})eV(t)\Upsilon_{21} \equiv
\frac{1}{2}\varepsilon_x f(t), $ and the diagonal elements are zero
for any order. Therefore, the effective Hamiltonian of our qubit is
$ H_e=\frac{1}{2}\varepsilon_z\sigma_z+\frac{1}{2}\varepsilon_x
f(t)\sigma_x, $ which is formally equivalent to the conventional ESR
Hamiltonian \cite{slichter}. For a sinusoidal perturbation
$f(t)=\cos \omega t$ at resonance ($\hbar\omega=\varepsilon_z$), the
time required for the $\pi$-operation, i$.$e$.$ $|G_+\rangle
\rightarrow |G_-\rangle$, is given by
\begin{equation}\label{eq:tpi}
t_\pi=\frac{2\pi\hbar}{\varepsilon_x}\sim \frac{\pi \hbar
\Delta_{2,1}}{\Upsilon_{12}^2
eV_0|E_{SL}|}(1-\frac{\varepsilon_{0z}^2}{\Delta_{2,1}^2}).
\end{equation}
\indent Since we hybridize the spin and orbital degrees of freedom,
orbital relaxation processes harm the spin coherence. We find that
acoustic phonon scattering is the dominant relaxation mechanism for
the energies relevant in our system. The electron-phonon scattering
Hamiltonian is
\begin{eqnarray}
H_{e-ph}&=&\sum_q\lambda_q(e^{iqz}b_q^\dagger+H.C.)\nonumber \\
&=&\sum_q(\Lambda_q^x\sigma_x+\Lambda_q^z\sigma_z)b_q^\dagger+H.C. \label{eq:projection} \\
&\equiv&\frac{1}{2}g\mu_B(B_x(t)\sigma_x+B_z(t)\sigma_z),
\end{eqnarray}
where $q$ is the phonon wave number, $\lambda_q$ is the coupling
constant and $b_q^\dagger$ is the phonon creation operator. In Eq$.$
\ref{eq:projection}, we project $H_{e-ph}$ to the qubit base, where
the effective coupling constants are $ \Lambda_q^x=\lambda_q \langle
1|e^{iqz}|2\rangle(C_{1,+}^{(1)}+C_{1,-}^{(1)})$ and $
\Lambda_q^z=\lambda_q \langle 1|e^{iqz}|3\rangle
(C_{1,+}^{(2)}-C_{1,-}^{(2)})$. $B_r(t)$ ($r=x,y,z$) represents the
fluctuating magnetic field caused by phonons. A standard Born-Markov
approximation \cite{slichter} gives relaxation ($T_1$) and coherence ($T_2$) times as follows,
\begin{eqnarray}
T_1^{-1} &=& k_{xx}(\varepsilon_z)+k_{yy}(\varepsilon_z), \\
T_{2phonon}^{-1} &=& (2T_1)^{-1}+k_{zz}(0),
\end{eqnarray}
where $k_{rr'}(\hbar\omega)\equiv \frac{1}{2}(g\mu_B)^2\int d\tau
\cos(\omega \tau)\langle B_r(t)B_{r'}(t+\tau)\rangle_0$ with
$\langle\cdots\rangle_0$ representing the thermal average. The
dephasing term $k_{zz}(0)$ related to $B_z(t)\sigma_z$ is negligible
since $|\frac{\Lambda_q^z}{\Lambda_q^x}|\sim\frac{\varepsilon_{0z}
M_{2,3}} {\Delta_{3,1}\Delta_{2,1}}\ll 1$ and is shown to be absent
in the zero frequency limit. This situation is similar to that of a
spin qubit with SO interaction \cite{khaetskii2001,golovach}. The
relaxation rate $T_1^{-1}$ is then
\begin{align}\label{eq:T1}
T_1^{-1} &= \frac{2\pi}{\hbar}\sum_q|\Lambda_q^x|^2
\delta(\varepsilon_z-\hbar \omega_q)\coth \frac{\beta
\varepsilon_z}{2} \nonumber \\
&\sim(\frac{E_{SL} \Upsilon_{1,2}
\Delta_{2,1}}{\Delta_{2,1}^2-\varepsilon_{0z}^2})^2\frac{1}{\tau_p(\varepsilon_z)},
\end{align}
with $\hbar\omega_q$ the phonon energy, $\beta=1/(k_B T)$, and the
relaxation time $\tau_p(E)$ defined in analogy with Fermi's Golden
rule for a transition from level 2 to 1 with energy transfer $E$.
The coherence time is obtained by
\begin{equation}
T_{2total}^{-1}=T_{2phonon}^{-1}+T_{2spin}^{-1},
\end{equation}
where we included the generic spin coherence time $T_{2spin}$, which
is the `pure' coherence time of the electron spin in the QD. The
upper bound of the quality factor of the one-qubit operation is
characterized by $2T_1$ divided by $t_{\pi}$ of Eq$.$
(\ref{eq:tpi}),
\begin{eqnarray}
Q\sim \frac{2\Delta_{2,1} \tau_p(\varepsilon_z)}{\pi\hbar}
\frac{eV_0}{|E_{SL}|}(1-\frac{\varepsilon_{0z}^2}{\Delta_{2,1}^2}).
\end{eqnarray}
Importantly, the quality factor is tunable by controlling the
amplitude of the ac voltage modulation, $V_0$.

\begin{figure}[htbp]
\includegraphics[width=8cm]{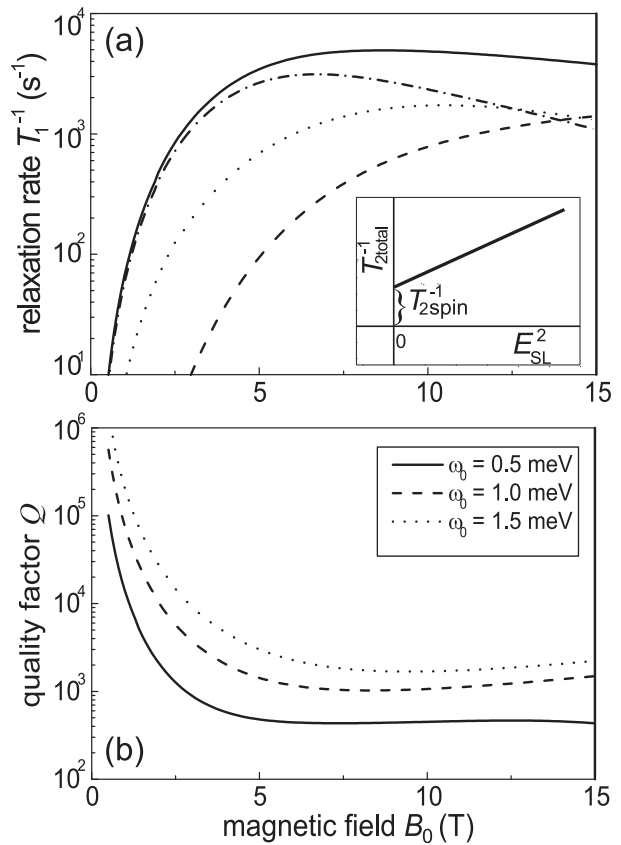}
\caption{\label{fig:rate-Q} (a) Relaxation rate $T_1^{-1}$ in a 1D
GaAs QD as function of external magnetic field $B_0$ due to
different phonon scattering mechanisms: deformation potential
(dashed), longitudinal piezoelectric (dotted), transversal
piezoelectric (dash-dotted). The solid line is the total scattering
rate. Inset:  schematic derivation of the pure electron spin
coherence time $T_{2spin}$ from the dependence of the total
coherence time $T_{2total}$ on the strength of the slanting field
$E_{SL}$. (b) $B_0$-dependence of the quality factor $Q$ for a
single qubit $\pi$ operation.}
\end{figure}

For the practical implementation of our scheme, 1D systems with
small electron-phonon coupling and/or weak SO coupling are
favorable. Single wall carbon nanotube QDs are very suitable,
because of the absence of piezoelectricity and the weak deformation
potential coupling \cite{suzuura}. SO coupling does not play a role
either. QDs in semiconductor (e$.$g$.$ SiGe) nanowires are also good
candidates, since 1D phonons couple weakly to the electron orbitals.
Here we estimate $T_1$ and $Q$ of GaAs 1D QDs embedded in bulk
AlGaAs. Since the phonon character is 3D in this system, the results
are worse than for the more suitable systems given above. Figure
\ref{fig:rate-Q}a shows the low-temperature ($\beta \varepsilon_z\gg
1$) relaxation rate caused by bulk acoustic phonons in a QD with
longitudinal parabolic confinement $\hbar\omega_0=1\ \mbox{meV}$ and
a transversal confinement of $10\ \mbox{meV}$. $E_{SL}=1\
\mu\mbox{eV}$, corresponding to $b_{SL}=1.16\ \mbox{T/$\mu$m}$,
which can be realized with a ferromagnetic material \cite{wrobel}.
Of the three acoustic phonon scattering mechanisms, transversal
piezoelectric scattering is dominant for low $B_0$, where $T_1$ is
of the order of 10 ms. For comparison, the typical relaxation time
from higher levels, i$.$e$.$ $|E_\pm\rangle\rightarrow
|G_\pm\rangle$ in Fig$.$~\ref{fig:level}b is much shorter, $\sim$10
ns, dominated by the deformation potential scattering. Note that the
contribution of SO interaction (Dresselhaus coupling) is very small
($T_{2SO}\sim 10^3 s$) in 1D QDs, in contrast to disk-shaped dots as
examined in Ref$.$ \cite{khaetskii2001}. Figure \ref{fig:rate-Q}b
shows $Q$ for various confinement potentials (dot length $L$) with
$b_{SL}=1.16\ \mbox{T/$\mu$m}$. $V_0=10\ \mu\mbox{V}$
\cite{wilfred,PAT}, and $t_{\pi}\sim 400$ ns for $\hbar\omega_0=1\
\mbox{meV}$. A quality factor $Q \gtrsim 10^4$ is often used as a
threshold for viable quantum computation
\cite{quantum-error-correction}.

We study the time evolution of the density matrix of the four levels
$|G_\pm\rangle, |E_\pm\rangle$, including $V_{ac}(t)$ and phonon
scattering. Near the resonant condition $\hbar \omega =
\varepsilon_z$, $|E_\pm\rangle$ are almost empty and do not
contribute to the qubit dynamics at all. $T_{2total}$ can be
evaluated using a time-resolved measurement of the Rabi oscillation
(see e$.$g$.$ \cite{engel0102}). After rotating the qubit over a
certain angle, a projection measurement is done into $|G_-\rangle$
or $|G_+\rangle$ using a single-shot read out scheme based on
spin-to-charge conversion \cite{elzerman}. Both energy selectivity
and spin selectivity of tunneling out of the QD are applicable for
our pseudo-spin qubit system. The read-out error introduced by level
mixing to $n=2$ by the slanting field is negligible, namely of the
order $(C_{1\sigma}^{(1)})^2\sim 10^{-6}$. Importantly, the pure
electron spin coherence time $T_{2spin}$ can be evaluated by
extrapolating the $T_{2total}$-dependence on $E_{SL}$ where
$T_{2phonon}^{-1}\propto E_{SL}^2$ (see Eq.~(\ref{eq:T1})), as shown
in the inset to Fig$.$~\ref{fig:rate-Q}a.\\
\indent For a universal set of quantum gates a two-qubit gate is
required. Here we present a realization of a two-qubit gate based on
two coupled dots in series \cite{Wiel03}. Although it has been
pointed out that an inhomogeneous magnetic field introduces swap
errors \cite{sousa2001,hu}, we show that correct swap operation is
possible in our system. The two-qubit Hamiltonian is $ {\cal
H}=\sum_{i=L,R}{\cal H}_{0i} +{\cal H}_T+{\cal H}_V, $ where ${\cal
H}_{0i}$ is the single-dot Hamiltonian $i=L,R$ (ac field is off,
$\varepsilon_x=0$), ${\cal H}_T$ represents the tunneling between
the dots, and ${\cal H}_V$ represents the inter-dot interaction $V$.
By projecting the Hamiltonian onto the qubits, we find
\begin{eqnarray}
{\cal H}_{0i}&=&\frac{\varepsilon_z}{2}\sum_\sigma \sigma c_{i\sigma}^\dagger c_{i\sigma}
+U n_{i\uparrow}n_{i\downarrow},\\
{\cal H}_T&=&\sum_\sigma[t_\sigma c_{L\sigma}^\dagger c_{R\sigma}
+s_\sigma\ c_{L\sigma}^\dagger c_{R-\sigma}+\mbox{H.c.}],\\
{\cal H}_V&=&V\sum_{\sigma\sigma'}n_{L\sigma}n_{R\sigma'}
\end{eqnarray}
where $c_{i\sigma}$ annihilates an electron of pseudo-spin $\sigma$
in dot $i$. A spin-dependent tunneling term $t_\sigma$ and a
tunneling term with spin flip $s_\sigma$ emerge, which are defined
by $ t_\sigma=C^{(0)2}_\sigma
t_{11}+C^{(1)L}_{1\sigma}C^{(1)R}_{1\sigma}t_{22}
+2C^{(2)}_{1\sigma}t_{13},
s_\sigma=(C^{(1)L}_{1\sigma}+C^{(1)R}_{1-\sigma})t_{12}, $ where
$t_{nm}$ represents the tunneling amplitude from level $n$ in dot
$L$ to level $m$ in dot $R$. The relevant lowest four eigenenergies
and their eigenfunctions are obtained by the effective exchange
Hamiltonian using local spin operators:
\begin{eqnarray}
{\cal H}_{EX}&=&J_{\parallel}S_{Lz}S_{Rz}+J_\perp(S_{Lx}S_{Rx}+S_{Ly}S_{Ry})\nonumber \\
&+&\tilde{\varepsilon}_z(S_{Lz}+S_{Rz}),
\end{eqnarray}
where $J_{\parallel}=\frac{2(t_\uparrow^2+t_\downarrow^2)}
{U-V}-\frac{4s^2(U-V)}{(U-V)^2-\varepsilon_z^2},
J_{\perp}=\frac{4t_\uparrow t_\downarrow} {U-V}$, and
$\tilde{\varepsilon}_z=\varepsilon_z(1-\frac{2s^2}{(U-V)^2-\varepsilon_z^2})$
with
$s=\frac{1}{2}(s_\uparrow+\frac{M_{1,2}^R}{M_{1,2}^L}s_\downarrow)$.
It is well known that the SO interaction makes the exchange
Hamiltonian anisotropic \cite{burkard}. In contrast to the SO case,
where the anti-symmetric term dominates, the dominant anisotropic
correction of ${\cal H}_{EX}$ in a slanting field is the symmetric
term. Nevertheless, C-NOT operation can be accomplished by this
anisotropic exchange Hamiltonian simply by replacing $J$ of the Heisenberg
Hamiltonian by $J_\parallel$, and single-qubit operation (SESR)
with replacing $\varepsilon_{0z}$ by $\tilde{\varepsilon}_z$,
as is shown in Refs$.$ \cite{burkard,hu}.\\
\indent In conclusion, we propose a viable qubit based on combining
the orbital and spin degrees of freedom of an electron in a QD
placed in a slanting Zeeman field. Both single-qubit rotation and
the C-NOT operation are demonstrated. This qubit is easier to
manipulate than a spin qubit and has a better quality factor than a
charge qubit. The concept is general and can be applied to a range
of systems such as single wall carbon nanotubes, GaAs, and SiGe QDs.
This scheme also allows for the measurement of the intrinsic single
electron spin coherence time. \\
\indent We thank Y$.$ Avishai, A$.$ Khaetskii and M$.$
Pioro-Ladriere for discussions. We acknowledge financial support
from DARPA grant number DAAD19-01-1-0659 of the QuIST program and SORST-JST.


\end{document}